\documentclass{ptapap}

\author{Ernst Paunzen}[UTFA]
\author{Monika Rode-Paunzen}[IFA]
\author{BEST}[BEST]
\affil[UTFA]{Department of Theoretical Physics and Astrophysics, Masaryk University \\
  Kotl\'a\v{r}sk\'a 2, 611\,37 Brno, Czech Republic}
\affil[IFA]{Department of Astrophysics, University of Vienna \\ 
T{\"u}rkenschanzstr. 17, A-1180 Wien, Austria}
\affil[BEST]{Bright Target Explorer (BRITE) Executive Science Team} 

\title{BRITE photometry of seven B-type stars}

\begin{document}

\maketitle

\begin{abstract}
We present the time series analysis of seven B-type stars observed with the BRITE satellites.
Furthermore, we located these stars in the Hertzsprung-Russell-diagram. One rotational induced
variable (HD 122980, $P$\,$\approx$\,80\,d) and one binary (HD 145502, $P$\,$\approx$\,5.6\,d)
showing variability due to ellipsoidal variation and/or reflection were detected.
\end{abstract}

\section{Introduction}

For B-type stars on the Main Sequence (MS) without any signs of activity,
we find pulsators, rotational and binarity induced variability. There are
$\beta$ Cephei stars with periods between three and eight hours of spectral 
type O9 to B3 which pulsate in low-order p- or g-modes \citep{Neilson2015}.
In addition, Slowly Pulsating B-type stars are objects which exhibit 
multi-periodic, high-order, and low-degree g-modes with periods of about 
one day \citep{Jerzykiewicz2013}.

The magnetic chemically peculiar (CP) stars of the upper MS are characterized by 
peculiar and often variable line strengths as well as photometric variability with 
the same periodicity and coincidence of extrema. Due to the chemical abundance 
concentrations at the magnetic poles, the show photometric variabilities correlated to
the rotational period \citep{Bernhard2015}. 

Binaries show all types of variability from eclipses, ellipsoidal or reflection variables,
just to mention a few. 

We investigated the BRITE light curves of seven B-type stars for variability. 

\begin{figure}[t]
\begin{center}
\includegraphics[width=\textwidth]{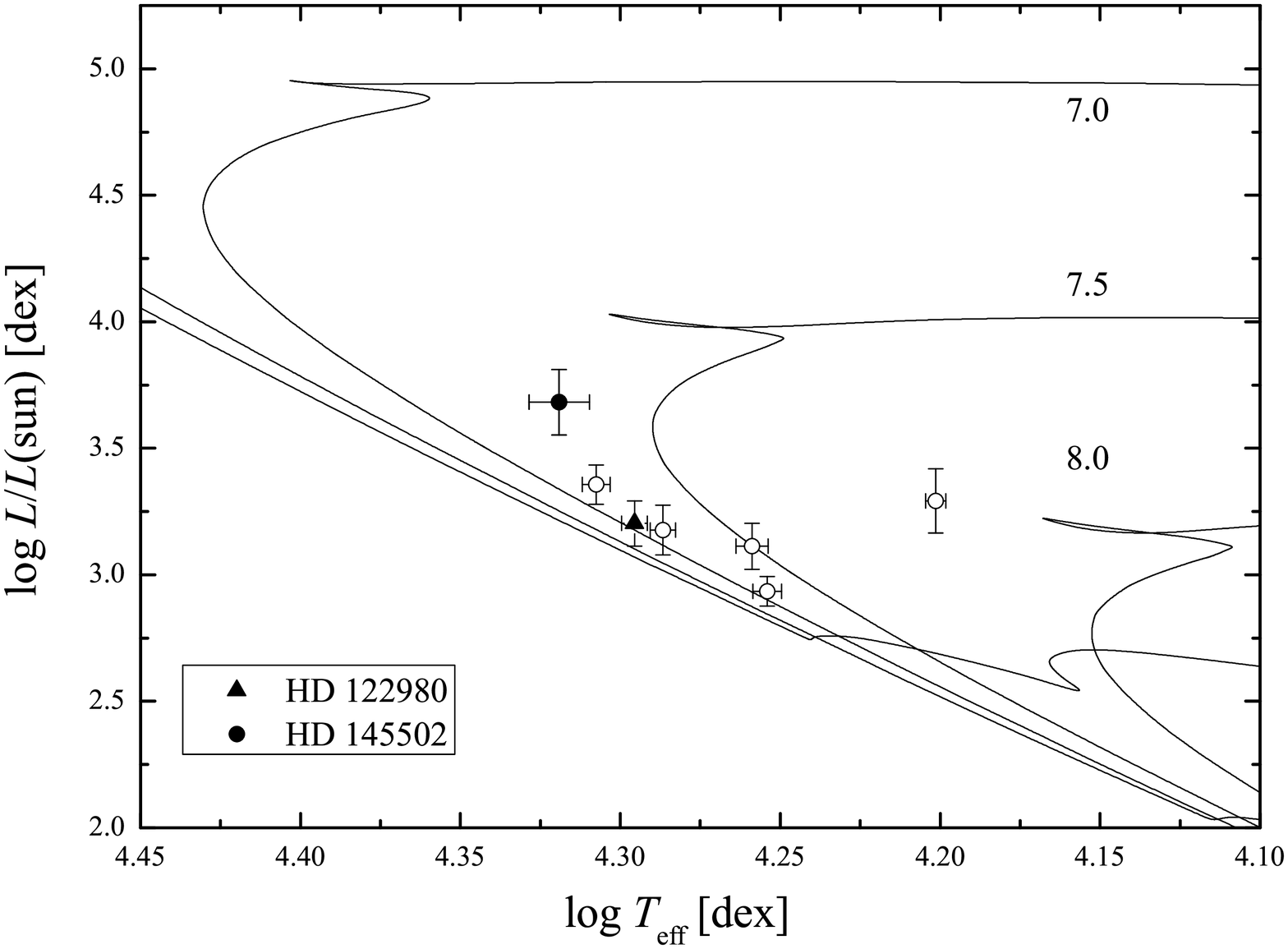}
\caption{The location of the targets stars (Table \ref{tab:data}) in a $\log L/L_{\odot}$ versus $\log T_\mathrm{eff}$
diagram together with the PARSEC isochrones \citep{Bressan2012} for solar metallicity ([Z]\,=\,0.019). HD 122980
and HD 145502 are the stars for which we detected variability.}
\label{fig:hrd}
\end{center}
\end{figure}

\section{Reduction and Data Analysis}

The reductions of the photometric data of the BRITE satellites
were carried out with a pipeline that takes into account bad pixels, 
median column counts, image motion, and PSF variations \citep{Popowicz2016}.
The photometry still needs to be decorrelated for several factors, for example
temperature variations within the satellites and detectors during
the orbit and on longer time scales, before
a scientific analyses of the intrinsic stellar variability can be done. We applied the
basic process of decorrelation as has been
described in detail by \citet{Pigulski2016}.

The data were obtained with four out of the five satellites:
BRITE-Austria (hereafter BAb), BRITE-Heweliusz (BHr), BRITE-Toronto (BTr),
and UniBRITE (UBr). The first satellite observes through the blue filter, the latter
three use the red one.

The light curves were examined in more detail using the PERIOD04 program \citep{Lenz2005}. 
The results were checked with those from the CLEANEST and Phase-Dispersion-Method computed 
within the programme package Peranso \citep{Paunzen2016AN}. The differences for the these
methods are within the derived errors depending on the time series characteristics, i.e. 
the distribution of the measurements over time and the photon noise.
If several data sets for one satellite were available, they were split and also separately analysed.

As next step, for the location of our programme stars in the Hertzsprung-Russell-diagram, 
the effective temperature ($T_{\rm eff}$) and luminosity ($\log L/L_{\odot}$) has to be derived.

First of all, the reddening (or total extinction $A_\mathrm{V}$) was estimated
using photometric calibrations in the Str{\"o}mgren-Crawford $uvby\beta$ \citep{crawford1978}
and the $Q$ parameter within the Johnson $UBV$ system \citep{Johnson1958}. 
These methods are only based on photometric indices and do not take
into account any distance estimates via parallax measurements. 
The agreement
is very good indicating that photometric measurements within both systems are intrinsically consistent.

The $T_{\rm eff}$ for the individual stars were calibrated by using the measurements in the
Johnson, Geneva 7-colour, and Str{\"o}mgren-Crawford photometric systems.
This approach was chosen because it guarantees the most homogeneous result. Taking $T_{\rm eff}$ values
from the literature, for example on the basis of spectroscopy, introduces all types of unknown biases. Those
are the usage of different stellar atmospheres, spectral resolution, analytic methods and so on. 
For the estimation of $T_\mathrm{eff}$, the following calibrations were used:
\begin{itemize}
\item Geneva 7-colour system: \citet{Cramer1984,Cramer1999,Kunzli1997}, based on the
reddening free parameters $X$ and $Y$ which are valid for spectral types
hotter than approximately A0. The results are therefore independent of the
estimation of $A_V$ for the program stars.
\item Johnson system: based on the $Q$ values for
luminosity class III and V objects according to the Tables listed by 
\citet{Schmidt1982}. The $(B-V)_0$ values for those luminosity classes 
were transformed into effective temperatures using the results by \citet[][Table 7]{Code1976}.
\item Str{\"o}mgren-Crawford system: \citet{Napiwotzki1993}, based on the
unreddened $[u-b]$ colour.
\end{itemize} 
The individual effective temperature values within each
photometric system were first checked
for their intrinsic consistency and then averaged. 

The absolute magnitude ($M_\mathrm{V}$) were directly calculated from the 
astrometric parallaxes ($\pi$) of the Hipparcos mission \citep{Leeuwen2007} using
the basic formula $M_\mathrm{V} = m_\mathrm{V} + 5\,({\rm log}\,\pi + 1) - A_\mathrm{V}$.

We used the $m_\mathrm{V}$ magnitudes as given by \citet{Kharchenko2001} who transformed 
the Tycho data to the Johnson system. This is the most
homogeneous available sample for this data type.
The bolometric corrections were taken from \citet{Flower1996} and 
an absolute bolometric magnitude of the Sun as  
$M_{\rm Bol}(\odot)$\,=\,4.75\,mag was used.
For the error estimate of $\log L/L_{\odot}$, only the error of the parallax was taken into account. 

The given spectral types are the best estimates from the extensive compilation by \citet{Skiff2014} whereas the
projected rotational velocities ($v \sin i$) are based on an updated version of the catalogue by \citet{Glebocki2000}.

\begin{figure}[t]
\begin{center}
\includegraphics[width=\textwidth]{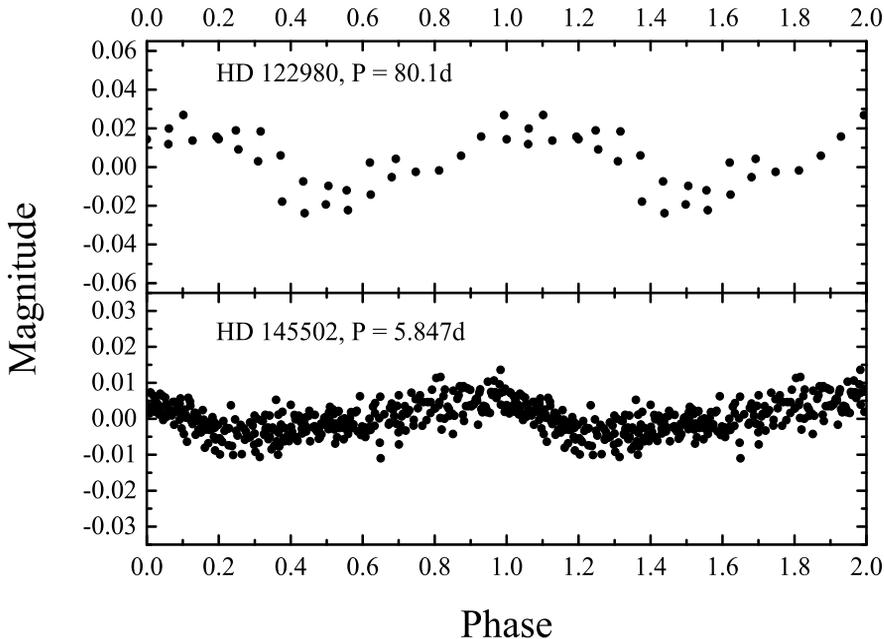}
\caption{The phase binned light curves for HD 122980 (5\,d binning of data) and HD 145502 (90\,min binning of data). 
HD 145502 is known spectroscopic binary system has an orbital period of 5.5521\,d (0.18\,c/d).}
\label{fig:phase}
\end{center}
\end{figure}

\begin{table}[t]
\begin{center}
\caption{Characteristics and astrophysical parameters of the targets.}
\label{tab:data}
\begin{tabular}{ccccccc}
\hline
\hline
HD & HIP & Spec. & $A_V$ & $\log T_\mathrm{eff}$ & $\log L/L_{\odot}$ & $v \sin i$ \\
& & & [mag] & [dex] & [dex] & [km\,s$^{-1}$] \\       
\hline
121790	&	68282	&	B2\,V	&	0.048	&	4.307(4)	&	3.36(8)	& 124 \\
122980	&	68862	&	B2\,V	&	0.064	&	4.296(4)	&	3.20(9)	&	18	\\
129116	&	71865	&	B3\,V	&	0.073	&	4.254(4)	&	2.93(6)	&		\\
133242	&	73807	& B5\,V	&	0.081	&	4.201(3)	&	3.29(13)	&	140	\\
144294	&	78918	&	B2.5\,V	&	0.071	&	4.259(5)	&	3.11(9)	&	295	\\
145482	&	79404	&	B2\,V	&	0.169	&	4.287(4)	&	3.18(10)	&	183	\\
145502	&	79374	& B2\,V	&	0.842	&	4.319(9)	&	3.68(13) & 155 \\
\hline
\end{tabular}
\end{center}
\end{table}

\section{Discussion}

The derived astrophysical parameters about the target stars is listed in Table \ref{tab:data}. 
The errors in the final digits of the corresponding quantity are given in parentheses.

Figure \ref{fig:hrd} shows the location of the targets stars (Table \ref{tab:data}) in a 
$\log L/L_{\odot}$ versus $\log T_\mathrm{eff}$
diagram together with the PARSEC isochrones \citep{Bressan2012} for solar metallicity ([Z]\,=\,0.019\,dex).
Only one star, HD 133242, is evolved on the Zero Age MS but still is well before the Terminal Age MS.
All other stars have ages $\log t$\,$<$\,7.5\,dex. In the following we discuss the individual objects in more
detail.

{\it HD 121790:} \citet{Telting2006} during their survey for $\beta$ Cephei pulsation of bright stars, found indications of 
spectroscopic variability with a line depth of 3\%, only. They published a $v \sin i$ value of 124 km\,s$^{-1}$. 
Only following upper limits for photometric variability were found: BAb: 2\,mmag; BTr: 0.9\,mmag, and UBr: 5\,mmag.
No signs of $\beta$ Cephei type pulsations could be detected within the available data sets.

{\it HD 122980:} with a $v \sin i$ value of only 18 km\,s$^{-1}$, one is tempted to think of a chemically peculiarity
for this star. If so, it would fall in the group of magnetic Helium peculiar stars \citep{Cidale2007}.
And indeed, \citet{Slettebak1968} found that the forbidden He\,{\sc i} 4469.9\AA\ line 
is very strong in HD 122980, but still this classification has to be confirmed. 
However, no magnetic field measurements are available, yet. For a B2\,V star (Table \ref{tab:data}),
seen equator-on, we estimate a rotational period of about 140\,d (0.007 c/d) which is the time base of the longest
data set (UBr). We therefore binned the data from 0.1 to 10\,d and searched for rotational induced variability.
Figure \ref{fig:phase} shows the phase binned data (5\,d) for a period of 80.1\,d (0.0125 c/d). The variation
is clearly visible implying $i$\,$\approx$\,35$^\mathrm{o}$. Furthermore,
\citet{Percy1974} listed a possible $\beta$ Cephei type
variability with a period of 0.035\,d (28.6 c/d). This period is very close to two times of 
the satellites orbital period and therefore very hard to detect. However, \citet{Balona1982} presented time series of five 
nights in Johnson $B$ and concluded that it is constant within a few thousands of magnitudes over several hours. 
We detected no signs of variability in this period domain with the following upper limits: BAb: 2\,mmag, 
BTr: 0.8\,mmag, and UBr: 5\,mmag. 

{\it HD 129116:} This star was often used as standard star for several photometric systems \citep{Strauss1981}. 
Up to now, no report
about apparent variability was published, yet. No significant frequencies were detected with upper limits of: 
BAb: 3\,mmag, BTr: 0.8\,mmag, and UBr: 4\,mmag.

{\it HD 133242:} A close binary system, resolved by speckle observations \citep{Horch2006}. No variations 
with an upper limit of 0.8\,mmag (BTr) were detected.

{\it HD 144294:} It has the highest $v \sin i$ value (295 km\,s$^{-1}$) among the sample. 
No variability was reported so far. The six
data sets show only upper limits between 0.9 to 4.6\,mmag, respectively.

{\it HD 145482:} This spectroscopic binary system was also investigated by \citet{Telting2006} who found no traces of variability. 
This result is confirmed with the available photometry. An upper limit of 1.2\,mmag (BHr) was deduced.

{\it HD 145502:} Only an upper limit (45 G) for a magnetic field was detected \citep{Bychkov2009} for 
this spectroscopic binary system.
The orbital period is determined by \citet{Levato1987} as 5.5521\,d (0.18\,c/d). The BLb data only yield upper 
limits between 30 and 60\,mmag,
but the UBr data clearly show a variation with a period between 0.17 and 0.18\,c/d. Figure \ref{fig:phase} shows 
the phase binned light curves for
one of the data sets. For the time series analysis, the original data were binned to a time resolution of 90\,min.
We assume that ellipsoidal variation and/or reflection are the causes of the photometric variability. Further
spectroscopic data are needed to further analyse the reasons of variations for HD 145502.

\acknowledgements{The paper is based on data collected by the 
BRITE Constellation satellite mission, designed, built, launched, operated and supported by the Austrian Research 
Promotion Agency (FFG), the University of Vienna, the Technical University of Graz, the Canadian Space Agency (CSA), 
the University of Toronto Institute for Aerospace Studies (UTIAS), the Foundation for Polish Science \& Technology 
(FNiTP MNiSW), and National Science Centre (NCN). }

\bibliographystyle{ptapap}
\bibliography{brite_paunzen}

\end{document}